# Synthetic CT Generation from MRI using 3D Transformer-based Denoising Diffusion Model

Shaoyan Pan[1,2] Elham Abouei[1], Jacob Wynne[1], Tonghe Wang[3], Richard L.J. Qiu[1], Yuheng Li[1], Chih-Wei Chang[1], Junbo Peng[1], Justin Roper[1], Pretesh Patel[1], David S. Yu[1], Hui Mao[4] and Xiaofeng Yang[1,2*]

[1]Department of Radiation Oncology and Winship Cancer Institute, Emory University, Atlanta, GA 30322, USA
[2]Department of Biomedical Informatics, Emory University, Atlanta, GA 30322, USA
[3]Department of Medical Physics, Memorial Sloan Kettering Cancer Center, New York, NY 10065, USA
[4]Department of Radiology and Imaging Sciences and Winship Cancer Institute, Atlanta, GA 30308

*Email: xiaofeng.yang@emory.edu

**Running title**: MRI-only Synthetic CT

**Abstract**

**Background:** Magnetic resonance imaging (MRI)-based synthetic computed tomography (sCT) simplifies radiation therapy treatment planning by eliminating the need for CT simulation and error-prone image registration, ultimately reducing patient radiation dose and setup uncertainty.

**Purpose**: In this work, we propose an MRI-to-CT transformer-based denoising diffusion probabilistic model (MC-DDPM) to transform MRI into high-quality sCT to facilitate radiation treatment planning.

**Methods:** MC-DDPM implements diffusion processes with a shifted-window transformer network to generate sCT from MRI. The proposed model consists of two processes: a forward process, which involves adding Gaussian noise to real CT scans to create noisy images, and a reverse process, in which a shifted-window transformer V-net (Swin-Vnet) denoises the noisy CT scans conditioned on the MRI from the same patient to produce noise-free CT scans. With an optimally trained Swin-Vnet, the reverse diffusion process was used to generate sCT scans matching MRI anatomy. We evaluated the proposed method by generating sCT from MRI on a brain dataset and a prostate dataset. Qualitative evaluation was performed using the mean absolute error (MAE) of Hounsfield unit (HU), peak signal-to-noise ratio (PSNR), multi-scale Structure Similarity index (MS-SSIM) and normalized cross correlation (NCC) indexes between ground truth CTs and sCTs.

**Results:** MC-DDPM generated brain sCTs with state-of-the-art quantitative results with MAE 43.317±4.104 HU, PSNR 27.046±0.817 dB, SSIM 0.965±0.005, and NCC 0.983±0.004. For the prostate dataset: MAE 59.953±12.462 HU, PSNR 26.920±2.429 dB, SSIM 0.849±0.041, and NCC 0.948±0.018. MC-DDPM shows statistically significant improvement in most metrics for both the brain and prostate sCTs relative to competing networks as evaluated using Student's paired t-test.

**Conclusions:** We have developed and validated a novel approach for generating CT images from routine MRIs using a transformer-based DDPM. This model effectively captures the complex relationship between CT and MRI images, allowing for robust and high-quality synthetic CT (sCT) images to be generated in a matter of minutes. This approach has the potential to greatly simplify the treatment planning process for radiation therapy by eliminating the need for additional CT scans, reducing the amount of time patients spend in treatment planning, and enhancing the accuracy of treatment delivery.

# 1. Introduction

Magnetic resonance imaging (MRI) and computed tomography (CT) are important imaging techniques in medical diagnosis and radiation therapy. In radiation therapy, both CT and MRI imaging are required for treatment planning. MRI provides anatomical and functional information with excellent soft tissue contrast while CT provides high geometrical accuracy and electron density information for the dose calculation in treatment planning [1,2]. MRI is advantageous over CT by providing superior soft tissue contrast near many tumor targets and more accurate organ at risk delineation [3-12] specifically in the pelvis, head and neck and brain. Complementary information from MRI and CT images is often necessary for accurate radiotherapy treatment planning.

The proposition of generating sCT images from MRI has garnered considerable attention. The adoption of MR-based sCT as a replacement for computed tomography (CT) has been suggested as a means of circumventing uncertainties associated with the co-registration of MRI and CT. It has also been suggested that this approach would mitigate the exposure to radiation inherent in CT imaging, the extra costs associated with multiple imaging modalities, and the discomfort experienced by patients. Promising applications of sCT includes treatment planning for radiation therapy as well as PET attenuation correction [9,13-15].

The techniques used to generate sCT images can be broadly classified into atlas-, segmentation-, sequence-, and hybrid-based methods[9]. Atlas-based methods utilize deformable image registration, but their effectiveness depends on the accuracy of the registration algorithm. Segmentation-based methods classify MRI into different types of tissue, such as soft tissue, air, and hard tissue, and assign a uniform electron density value to each type. The accuracy of this method is contingent upon the segmentation algorithm, but it may not be very reliable due to the similarities between soft tissue and bone in MRI. Sequence-based methods employ two or more specialized MRI sequences to obtain different types of MRI, which can increase scanning time, motion artifacts, and patient discomfort. Hybrid methods are a combination of two or more of the above techniques used to generate sCT images.

Artificial intelligence and machine learning techniques have advanced rapidly in recent years. These techniques can be utilized for image synthesis and are broadly categorized into dictionary learning-, random forest-, and deep learning-based models. Dictionary learning-based methods are sensitive to MR intensities that can vary based on scanning parameters and tissue types. Deep learning-based image synthesis methods rely on complex, nonlinear, and trainable mapping methods, such as convolutional neural networks (CNNs)[16] and generative adversarial networks (GANs)[17] GAN-based approaches have been particularly successful in generating realistic CT images from MRI. Wolterink *et al.*[18] use GAN with unpaired brain MR and CT images. Lei *et al.*[19] developed sCTs from MRIs based on dense cycle GAN model to effectively capture the relationship between the CT and MRIs. Their method generated robust, high-quality sCT in minutes on brain and pelvic datasets. Zhao et al.[7] proposed an approach using a hybrid CNN and transformer architecture as a generator in the GAN framework. Their results show that the proposed method can generate accurate CT images from pelvic MRI and is robust against local mismatch between MR and CT images. However, these methods all suffer from unstable training, mode collapse, and output homogeneity due to the adversarial method used to train GANs, as has been previously described[20,21].

As an alternative to GAN, diffusion and score matching models are generative approaches inspired by non-equilibrium thermodynamics in physics. They define a Markov chain of diffusion steps to gradually add random noise to data and then learn to reverse the diffusion process to generate samples from the noise[22-26]. Diffusion models utilize a neural network (typically a U-shape CNN) to learn denoising. As opposed to GANs, diffusion models are not reliant upon adversarial training methods[27]. This improves training stability and results in more authentic output images with higher quality and greater semantic diversity. Several diffusion-based generative models[1,28-31] have been proposed for medical image synthesis and demonstrate state-of-the-art image quality superior to CNN-based and GAN-based methods.

Our study presents a novel 3D diffusion-based approach, MRI-to-CT denoising diffusion model (MC-DDPM) for generating 3D sCT from MRI images. We trained our model on paired MRI-CT data, where MRI was used as the condition and CT was the target. By gradually translating standard Gaussian noise to the target CT conditioned on the MRI counterpart, our trained model can generate high-quality sCT with reduced artifacts. Our experiments on brain and prostate T1-weighted MRI-CT datasets demonstrate the effectiveness of our proposed method, enabling high-quality sCT generation. This is the first 3D DDPM method for MRI-to-CT synthesis.

Strengths:

- Image quality: MC-DDPM better captures the complex structures of high-dimensional data to generate high-quality sCTs. This is because the MC-DDPM uses a sequential process of diffusion steps to generate samples, which allows for more precise control by the neural network over the generated samples.
- Stability: MC-DDPM are more stable than GANs during training, which means they are less prone to mode collapse and less affected by the hyper-parameters. This is because the MC-DDPM uses a diffusion process to gradually generate samples, which reduces the likelihood of generating unrealistic or inconsistent samples.

Limitations:

- MC-DDPM is computationally expensive due to its reliance on a long Markov chain of diffusion steps to generate samples.

## 2. Method

The proposed MC-DDPM translates a patient's MRI into a sCT image. As shown in the Fig. 1, a diffusion-based process converts a three-dimensional isotropic Gaussian noise $T \sim \mathcal{N}(0,1)$ into a sCT image conditioning on the corresponding MR. The diffusion process relies on a significant assumption: by adding a small amount of noise $\epsilon$ to a real CT scan $X$ over n timesteps, we can transform $X$ into a purely Gaussian noise sample T, with n being sufficiently large. Consequently, the noise-free image X can be retrieved by eliminating the added noise from the noise sample T, which can also be generalized to any noise sample. Consequently, the diffusion process consists of two processes: a forward process that adds Gaussian noise to the CT image, and a reverse process learned by the proposed neural network, Swin-Vnet, that removes the Gaussian noise from the noisy CT images. However, in practice, the reverse process is stochastic, resulting in unpredictable but realistic CT scans that cannot be specifically attributed to a particular patient subject. To overcome this limitation, an additional MR scan is used to guide the diffusion process and generate paired CT images for the same patient.

The Swin-Vnet architecture is a three-dimensional encoder-decoder based on a Shifted-window (Swin) vision transformer. The encoder performs downsampling on the input data $X_t$, which is a concatenation of the noisy CT scan $Y_t$ at timestep t and the MR scan from the same patient and generates compressed features using one early residual convolutional block and three Swin-transformer-based blocks. The decoder, which has a symmetrical structure to the encoder, performs upsampling and decompresses the features to estimate the less noisy CT scan $Y_{t-1}$ at timestep t. The architecture of MC-DDPM and the mathematical formulation of the diffusion process are introduced here.

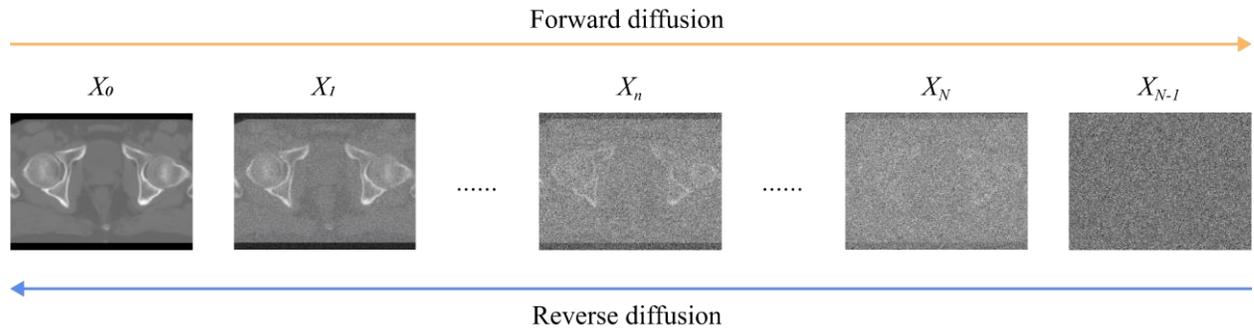

**Figure 1:** The proposed diffusion process of the MC-DDPM's synthesis: First, the 3D CT image is transformed into pure Gaussian noise through forward diffusion, which involves iterative addition of a small amount of noise. Second, to obtain the noise-free image, a neural network is used to repeatedly denoise the Gaussian noise through a reverse process.

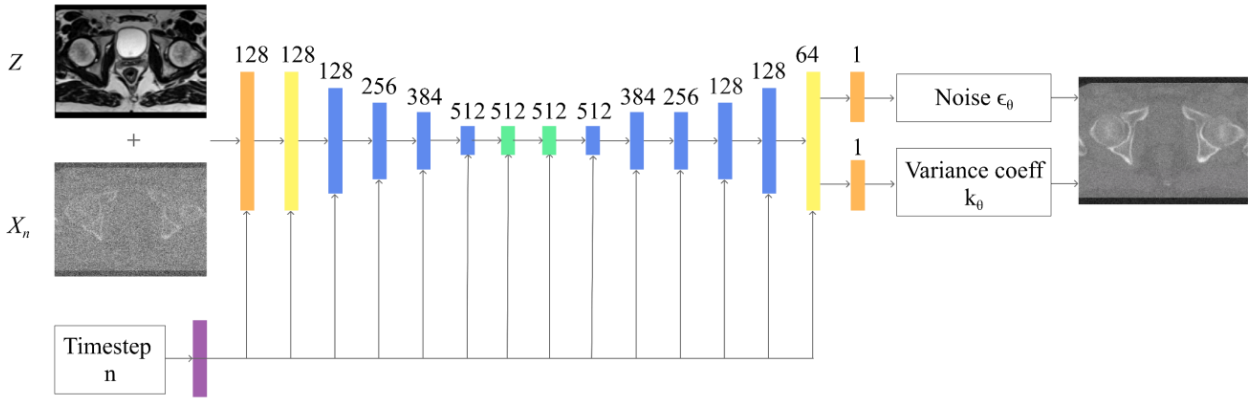

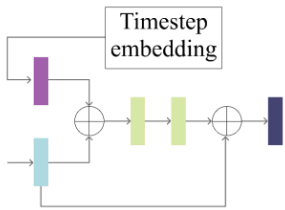 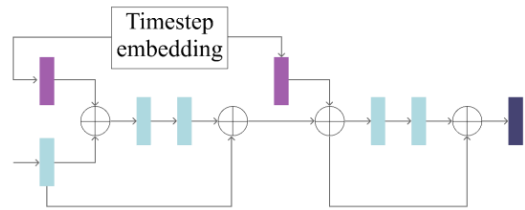

**Figure 2:** a) Network architectures of the Swin-Vnet used in the reverse diffusion process for the prostate MRI-to-CT synthesis: a symmetrical encoder-decoder architecture is utilized to predict the noise and variance coefficient which are used for the denoised CT images. Notice that the Swin-attention middle blocks are the Swin-attention blocks without the last up/down-sampled layer. b) The Swin-transformer block consists of a window self-attention and a shifted window self-attention module. c) In the convolutional block, two convolutional residual structures are used to learn local features.

## 2.A Network architecture

Fig. 2 illustrates an overview of the architecture for the prostate MRI-to-CT synthesis (the detailed network architecture for brain MRI-to-CT synthesis is shown in Appendix. A). The input medical images are initially processed through a 1×1 convolutional layer with stride of 1 in the encoder to learn the early features. Our proposed encoder architecture begins with one down-sampled convolutional block that learns the local characteristics from the high-resolution inputs. This step is followed by four sequential down-sampled Swin-attention blocks that learn global information from the low-resolution features. Two middle Swin-attention blocks, without down-sampling or up-sampling, are then connected to further compute the global characteristics. The decoder comprises four up-sampled Swin-attention blocks and one up-sampled convolutional block that restore the features to their original resolution. Following the processing through the encoder and decoder, the resulting features are passed through a final convolutional layer with a kernel size of 3x3 and stride of 1. This network estimates the noise and variance interpolation coefficient, which are subsequently used to calculate the less noisy CT scans at timestep n-1. Further details are provided in Section 2.B.3. To supply the network with the relevant temporal information, the timestep is encoded using sinusoidal embedding (SE)[24], with a maximum period of $10^6$ and a feature dimension of 128. These timestep embeddings are then fed into all blocks within the network, ensuring that the network is aware of the precise timestep at which each image should be denoised. Moreover, we applied the residual connection[32] in both the Swin-attention blocks and convolution blocks to improve network stability and reduce overfitting. Furthermore, as illustrated in Figure b), we incorporate a shortcut connection across the blocks via matrix concatenation, linking each encoder block to a decoder block in the same resolution level. This connection effectively conveys high-resolution information from the encoder to the decoder, thereby enhancing the accuracy of the estimation process.

### 2.A.1 Convolutional block

The up/down-sampling convolutional blocks in our architecture consist of a timestep expanding (linear) layer, four subsequent convolutional layers, and a final up/down-sample layer using bilinear interpolation. The convolutional layers have 3×3 spatial filters with stride of 1 in all axes, and the last section utilizes a trilinear up/down-sampling interpolation before a convolutional layer (kernel size $1 \times 1$). The timestep expanding layer re-embeds the timestep embeddings to align with the convolutional layer's dimension. More specifically, the timestep expanding layer takes an input timestep embedding $T \in \mathbb{R}^{1 \times d}$ and outputs a scale feature map $sc \in \mathbb{R}^{1 \times d_c}$ and a shift map $sh \in \mathbb{R}^{1 \times d_c}$, where $d$ is the timestep dimension and $d_c$ is the convolutional feature map dimension. The outputs of the first convolutional layer $h$ and the scale and shift feature map are then summed via a scale shift normalization [24]:

$$O = h * (1 + sc) + sh \qquad [1]$$

The output is then passed to subsequent convolutional layers, allowing each convolutional section to learn semantic information conditioned on timestep t. A residual connection is applied between the early convolutional layer and the final output, while group normalization[33] (with 32 groups) and SiLu[34] activation functions are applied to each of the convolutional layer.

### 2.A.2 Swin-attention block

Each Swin-attention block is composed of a window self-attention (W-SA) module, a shifted-window self-attention (SW-SA) module as depicted in Fig. 1 (b), and a trilinear up/down-sampling interpolation before a convolutional layer with kernel size of $1 \times 1$. The W-SA module employs a window-partitioning layer to divide the input features into non-overlapping windows, followed by a multi-head self-attention (MHSA) layer to compute the global information for each window. A linear layer is then used to embed all the windows into a feature map. The SW-SA module has the same structure as the W-SA module, except that a predetermined distance is used to shift the non-overlapping windows.

Given the input features $F \in R^{H \times W \times L \times C}$, in the W-SA module, we first partition the embedded feature into non-overlapping windows using the window partition layer. Specifically, $F_e$ is divided into $\frac{H}{N} \times \frac{W}{N} \times \frac{L}{N_L} \times C$ windows with the size of $N \times N \times N_L$, where N is empirically set to 4, 4, 4, 2 for the first to the fourth Swin transformer blocks, and $N_L$ is set to 4, 2, 2, 2, respectively. For each window $W$, we employ a MHSA module that consists of $P$ parallel self-attention heads, each of which learns global features across the window. Each head comprises independent weight matrices Query (Q), Key (K), and Value (V), which performs:

$$head_p = softmax(\frac{W_l Q_p (W_l K_p)^T}{\sqrt{d_k}})(W_l V_p) \quad [2]$$

$$A_l = concat(head_p, \ldots, head_P) W_o \quad [3]$$

where $A_l \in R^{(N \times N \times N_L)}$ is the attentions calculated for the l'th window, and $W_o$ is another weight matrix. By gathering all $A_l$, we obtain an attention map $F_o$ with the same size as the input W. An additional linear layer is connected to $F_o$ for better embedding performance. $F_o$ is then reshaped back to the original size of the non-partitioned features $F$ as the final output $F_e^{out} \in R^{H \times W \times L \times C}$.

$X_e^{out}$ is then passed into the SW-SA module, which shifts the partitioned windows by $(\frac{N}{2}, \frac{N}{2}, \frac{N_L}{2})$ voxels. We thus obtain attention information not only from W-SA modules, but also between windows. In summary, a set of W-SA and SW-SA modules performs:

$$F_e^{out} = linear(W - SA(F_e)) \quad [4]$$

$$S_{out} = linear(SW - SA(F_e^{out})) \quad [5]$$

where *linear* is linear layer. Group normalization (with 32 groups) and SiLu activation are applied after all W-SA, SW-SA modules, and the linear layers.

**2.B Diffusion process**

We present a three-stage formulation for the proposed diffusion model: first, a forward diffusion process is executed, wherein small amounts of Gaussian noise $T$ are gradually applied to a given CT image $X_0$ over $N$ timesteps, to gradually transform the CT scan into pure multi-dimensional Gaussian noise $X_N$. Next, the proposed MC-DDPM network is trained to learn a reverse diffusion process, conditioned on the MRI $Z$, which effectively removes the small amounts of noise overlayed at each timestep, and subsequently denoises $X_N$ back to the original image $X_0$. With an optimal MC-DDPM denoiser, we can recursively remove Gaussian noise to obtain the noise-free CT image paired to the input MRI.

### 2.B.1 Forward diffusion

In the forward diffusion process, the noisy image at timestep n is defined to depend solely on the noisy image at timestep n-1. Specifically, we define the noisy image generation process $q$ as a Markov process where the transition probability from image $X_{n-1}$ to $n$ follows a Gaussian distribution $\mathcal{N}$:

$$q(X_n|X_{n-1}) = \mathcal{N}(X_n; \sqrt{1-\beta_n}X_{n-1}, \beta_n I) \tag{6}$$

where $\beta_n$ is a pre-determined variance at timestep $n$. Practically, we are able to efficiently represent noisy images at any arbitrary timestep $n$ using reparameterization[23]:

$$X_n = \sqrt{\prod_{i=1}^{n}(1-\beta_i)}X_0 + \sqrt{1-\prod_{i=1}^{n}(1-\beta_i)}\epsilon_n) \tag{7}$$

where $\epsilon_n \sim \mathcal{N}(0, I)$ is noise sampled from a normal distribution. The selection of the maximum timestep $N$ is 1000 and $\beta_n$ is $5e^{-6}n$ for all experiments.

### 2.B.2 Reverse diffusion

In the reverse diffusion process, we calculate the inverse Gaussian distribution $p(X_{n-1}|X_n) = \mathcal{N}(X_{n-1}; \mu_n, \Sigma_n)$ so we can recursively move in a reverse direction from $X_T$ to $X_0$. Following Ho et al. and Song et al.'s works[26], the inverse probability can be calculated in a closed form only when the clean CT scan $X_0$ is already known,

$$\mu_n = \frac{\beta_n \sqrt{\prod_{i=1}^{n-1} 1-\beta_i}}{1-\prod_{i=1}^{n} 1-\beta_i} X_0 + \frac{\sqrt{1-\beta_n}(1-\prod_{i=1}^{n-1} 1-\beta_i)}{1-\prod_{i=1}^{n} 1-\beta_i} X_n \tag{8}$$

$$\Sigma_n = \frac{1-\prod_{i=1}^{n-1} 1-\beta_i}{1-\prod_{i=1}^{n} 1-\beta_i} \beta_n \tag{9}$$

which is impractical since the clean scan should be unknown for the testing data. Accordingly, we propose a neural network $\theta$ to approximate the $p(X_{n-1}|X_n)$:

$$p_\theta(X_{n-1}|X_n, Z) = \mathcal{N}(X_{n-1}; \mu_\theta(X_n, n|Z), \Sigma_\theta(X_n, n|z)) \tag{10}$$

where $Z$ is the conditioning MRI, $\mu_\theta$ is an estimated mean matrix and $\Sigma_\theta$ is an estimated variance matrix of the inverse Gaussian distribution. Therefore, the ground truth $\mu_n$ and $\Sigma_n$ are served as optimization targets for the estimated $\mu_\theta$ and $\Sigma_\theta$. Using reparameterization, the less noisy image $X_{n-1}$ can be calculated as:

$$X_{n-1} = \mu_\theta(X_n, n|Z) + \sigma_\theta(X_n, n|Z) * \epsilon \tag{11}$$

where $\epsilon \sim \mathcal{N}(0, I)$ is a noise sampled from a normal distribution, $\sigma_\theta$ is the standard deviation of the inverse distribution. In conclusion, we apply a transformer-based network (MC-DDPM network) taking inputs of a noisy image $X_n$, timestep $n$, and MR scan $Z$ from the same patient, to estimate the mean and standard deviation of the inverse distribution. We thus generate the less noisy image $X_{n-1}$ without foreknowledge of the clean image $X_0$.

## 2.B.2.a) Estimating the mean

It may appear that the most straightforward approach to estimating the mean µ using MC-DDPM may be to estimate it directly; however, this can be a challenging task since the value of the mean is unconstrained [23], leading to unstable training which may potentially reduce the quality of the generated synthetic images. To enhance the stability of the network and improve the quality of the synthetic images, we optimize MC-DDPM as a noise predictor rather than a direct mean predictor. As demonstrated in [23], the mean $\mu_\theta$ can be calculated using the estimated noise $\epsilon_\theta$:

$$\mu_\theta(X_n, n|Z) = \frac{1}{\sqrt{1-\beta_n}}\left(X_n - \frac{\beta_n}{\sqrt{1-\sum_{i=1}^{n} 1-\beta_i}}\epsilon_\theta(X_n, n|Z)\right) \quad [12]$$

where the noise $\epsilon_\theta(X_n, n|Z)$ is the only unknown parameter. MC-DDPM is therefore optimized to predict the noise. The noise prediction is therefore formally equivalent to minimizing the mean absolute error (MAE) between the predicted noise $\epsilon_\theta$ at the non-resampled timestep $n$ and the ground truth noise $\epsilon_n$ over all possible timesteps $n$:

$$\underset{\epsilon_\theta}{\mathrm{argmin}}\ L_{mean} = MAE(\epsilon_n, \epsilon_\theta(X_n, n|Z)) = E_v(E_n(||\epsilon_n - \epsilon_\theta(X_n, n|Z)||_1^1)) \quad [13]$$

where $E_n$ is the average loss over all timestep $n$, and $E_v$ is the average over all voxels.

## 2.B.2.b) Estimating variance

The variance can be obtained as equal to the pre-determined variance $\beta_n$ in the forward process [23], or directly estimated by the MC-DDPM network itself. In this work, we employ Prafulla et al.'s work [24] to first train the network to estimate an interpolation variance coefficient $k_n$ at timestep $n$ which can be used for the variance matrix:

$$\Sigma_\theta(X_n, n|Z) = \exp\left(\frac{k_\theta(X_n, n|Z)+1}{2} * \log\beta_n + \frac{(1-k_\theta(X_n, n|Z))}{2} * \log\left(\frac{1-\prod_{i=1}^{n-1}(1-\beta_i)}{1-\prod_{i=1}^{n}(1-\beta_i)}\beta_n\right)\right) \quad [14]$$

where $k_\theta(X_n, n|Z)$ is only unknown parameter. To obtain the coefficient $k_\theta$, the network is optimized by a variational low bound loss:

$$\underset{k_\theta}{\mathrm{argmin}}\ L_{var} = E_v(E_n(L_{VLB}(X_n, n))) \quad [15]$$

where $L_{VLB}$ is defined motivated by Prafulla et al.'s [24] and Ho et al.'s[35] implementations with modified boundary conditions $L_{high}$ and $L_{low}$:

$$L_{VLB}(X_n, n) = \begin{cases} \frac{0.5}{\log 2} L_n, & n > 0 \\ -\frac{1}{\log 2} \log(L_{high} - L_{low}), & n = 0 \text{ and } \frac{1}{D} - 1 < X_n < 1 - \frac{1}{D} \\ -\frac{1}{\log 2} \log(1 - L_{low}), & n = 0 \text{ and } 1 > X_n > 1 - \frac{1}{D} \\ -\frac{1}{\log 2} \log(L_{high}), & n = 0 \text{ and } -1 < X_n < \frac{1}{D} - 1 \end{cases} \quad [16]$$

$$L_n := \log \Sigma_\theta(X_n, n|Z) - \log \Sigma_n + \frac{\Sigma_n}{\Sigma_\theta(X_n, n|Z)} + \frac{(\mu_\theta(X_n, n|Z) - \mu_n)^2}{\Sigma_\theta(X_n, n|Z)} - 1, \quad [17]$$

$$L_{high} := GELU\left((X_n - \mu_\theta(X_n, n|Z) + \frac{1}{D}) * \exp(-0.5 \log(\Sigma_\theta(X_n, n|Z)))\right), \quad [18]$$

$$L_{low} := GELU\left((X_n - \mu_\theta(X_n, n|Z) - \frac{1}{D}) * \exp(-0.5 \log(\Sigma_\theta(X_n, n|Z)))\right), \quad [19]$$

where $D$ is the number of possible intensities of the image, which is set to (1024+1650, referred to the clipping range described in Section. 3) since we propose the diffusion process for CT volumes. The $GELU$ (Gaussian Error Linear Units function) is adopted as a fast approximation of the Cumulative Distribution Function for Gaussian Distributions[24]. Notice that there is only one parameter to be optimized, which is the $k_\theta(X_n, n|Z)$ in $\Sigma_\theta(X_n, n|Z)$.

The overall optimization function is presented as:

$$L = L_{mean} + \gamma L_{var} \quad [20]$$

where $\gamma$ is a weighting parameter which is empirically selected as the ratio between the inference diffusion step to the training diffusion step.

### 2.B.3 Training the diffusion process with resampled steps

In the training stage, we randomly select a batch of timesteps $n$ from the training timestep set $n \in [1, 2, 3, ..., N]$ (the maximum timestep $N$ is empirically selected as 1000), to generate noisy CT scans. Accordingly, we were expected to use the finely trained network to generate clean scans from Gaussian noise using the same timesteps, which is inefficient. Therefore, we almost evenly spaced numbers between 1 and 1000 by 50 timesteps and denote the new number set as the resampled timestep set $s \in [s_1, s_2, ..., s_J] = [1, 21, 41, ..., 1000]$. Then the network is optimized to estimate the noise and variance coefficient at the corresponding resampled timesteps $s$.

### 2.B.4 Generated synthetic CT volume

Given a new noisy Gaussian sample $X_s$, the MC-DDPM can generate a less noisy image $X_{s-1}$ by Eqn. (9), (10) and (12) and all we need is to provide the MR scan $Z$ and a new gaussian noise $\epsilon_{s_j} \sim \mathcal{N}(0, I)$:

$$X_{s_{j-1}} = \frac{1}{\sqrt{1-\beta_{s_j}}} \left( X_{s_j} - \frac{\beta_{s_j}}{\sqrt{1-\sum_{i=1}^{s_j} 1-\beta_i}} \epsilon_\theta \left( X_{s_j}, s_j | Z \right) \right) +$$

$$\sqrt{\exp\left( \frac{k_\theta(X_{s_j}, s_j|Z)+1}{2} * \log \beta_{s_j} + \frac{1-k_\theta(X_{s_j}, s_j|Z)}{2} * \log\left( \frac{1-\prod_{i=1}^{s_j-1}(1-\beta_i)}{1-\prod_{i=1}^{s_j}(1-\beta_i)} \beta_{s_j} \right) \right)} * \epsilon_{s_j} \quad [19]$$

We recursively denoise $X_s$ until we obtain the final CT image $X_0$, a synthetic image anatomically matching the input MRI. Notice that the noise $\epsilon_{s_j}$ introduces randomness into the generation process, we generate the final sCT scan in a manner of Monte Carlo-based (MC-based) generation: we run the generation process 5 times for each patient and took the average result as the final sCT scans.

## 3. Data Acquisition and Preprocessing

### 3.A Institutional brain dataset

We utilized an additional dataset for our MRI-to-sCT translation, which consists of MR-CT scan pairs from 36 patients with brain imaging. The MR scans were obtained from Siemens Avanto MRI scanner, while the CT scans were collected by Siemens SOMATOM Definition AS CT scanner. To align the MR scans with their corresponding CT scans, we used Velocity AI 3.2.1 which is implemented for clinic use in our department. Each MR and CT scan has a voxel size of 1x1x1 mm3, with 234~286 slices in each scan. Each slice has 512x512 pixels. We centered, removed the air background, and central-cropped the slices to highlight the Cerebrospinal fluid region. The final MR-CT scan pairs have a voxel size of 192x192x96. For training and inference, we used the same patch-based input approach and sliding window prediction technique as with the institutional prostate dataset. The patch size chosen for the brain dataset was 64x64x4. To generate sCT images during inference, we use a sliding window approach with a window size set equal to the patch size. There is a 50% overlap of the patch size, with Gaussian weighting applied to the edges of the windows. No augmentation was used during this process. Prior to training and inference, the voxel intensities of CT scans were cut to [-1024, 1650], and jointly normalized to the interval [-1, 1]. The voxel intensities of MRIs were independently normalized to [-1, 1]. We trained our model using the first 28 patients in the dataset, while the remaining 6 patients were used for testing. Two additional patient samples were reserved for validation. MC-DDPM was also trained using an AdamW optimizer with an initial learning rate of 3e-5 and weight decay of 1e-5 across 500 epochs.

### 3.B Institutional prostate dataset

Our goal is to create synthetic CT images (sCT) from MRI scans, utilizing a dataset that consists of MRI and CT scans collected from 28 patients. The MRIs were obtained from GE Signa MRI scanner, while the

CT scans were acquired by Siemens SOMATOM Definition AS CT scanner. To align the MRIs with the CT scans from the corresponding patient, we used Velocity AI 3.2.1 which is implemented for clinic use in our department. Each CT scan and post-registered MRI consists of 341~477 slices, with each slice having 512x512 pixels. The voxel spacing is 1x1x1 millimeters$^3$ (mm$^3$). With the guidance of two expert physicians, we focused on the anatomy structure containing bladder and prostate in each scan, centering and cropping each MR and CT scan in the sagittal, coronal, and axial directions to minimize irrelevant background. The final MR-CT pairs have a voxel size of 256x256x32. During each training stage, two patches with size 128×128×4 are randomly selected from an MR-CT pair. Normalization techniques, and inference strategies used for the institutional brain dataset were also applied to this prostate dataset.

For training, we used the first 20 patients in our dataset, while another 6 patients were used for testing. The remaining 2 patients were used for validation. The MC-DDPM was trained using a AdamW optimizer[36] with an initial learning rate of $10^{-4}$ and weight decay of $3 \times 10^{-5}$ across 800 epochs, where each epoch indicates a complete iteration through the entire training dataset.

## 4. Implementation and performance evaluation

### 4.A Implementation details

All experiments were conducted using the PyTorch framework in Python 3.8.11 running on a Windows 11 workstation equipped with a single NVIDIA RTX 6000 GPU with 48GB memory.

### 4.B Synthetic CT Evaluation

To measure the quality of the sCT from the proposed MC-DDPM, we compare the sCT and the clinical CT. We evaluate the mean absolute error (MAE) of Hounsfield unit (HU), peak signal-to-noise ratio (PSNR) of decibel (dB), multi-scale structure similarity index (MS-SSIM) with evaluation scale of 5 and normalized cross correlation (NCC) indices to quantify the absolute difference, peak signal similarity, image overall visual similarity, and image correlation, respectively. Greater PSNR, MS-SSIM (range from 0 to 1) and NCC (range from 0 to 1) indicate better quality of the sCT. Final performance was reported after evaluating all sCT scans. We compared MC-DDPM's performance to state-of-the-art methods, including the MRI-to-CT pixel-to-pixel generative adversarial network (MR-GAN)[19], MRI-to-CT cycle-consistent generative adversarial network (MR-CGAN)[19], 2D improved DDPM (2D-IDDPM)[25], and 3D DDPM (3D-DDPM)[24]. To ensure a fair comparison, we configured the competing networks according to their corresponding references. For 3D-DDPM, we extended the reported 2D configurations to three dimensions, maintaining the same number of layers, the same number of filters in each layer. We also used identical training hyperparameters for all networks, including patch size, optimizer, learning rate, training and generation timesteps (only applicable for DDPM-based methods), and data preprocessing. Pair-wise comparisons between MC-DDPM and competing networks were made using Student's paired t-test with α =0.05.

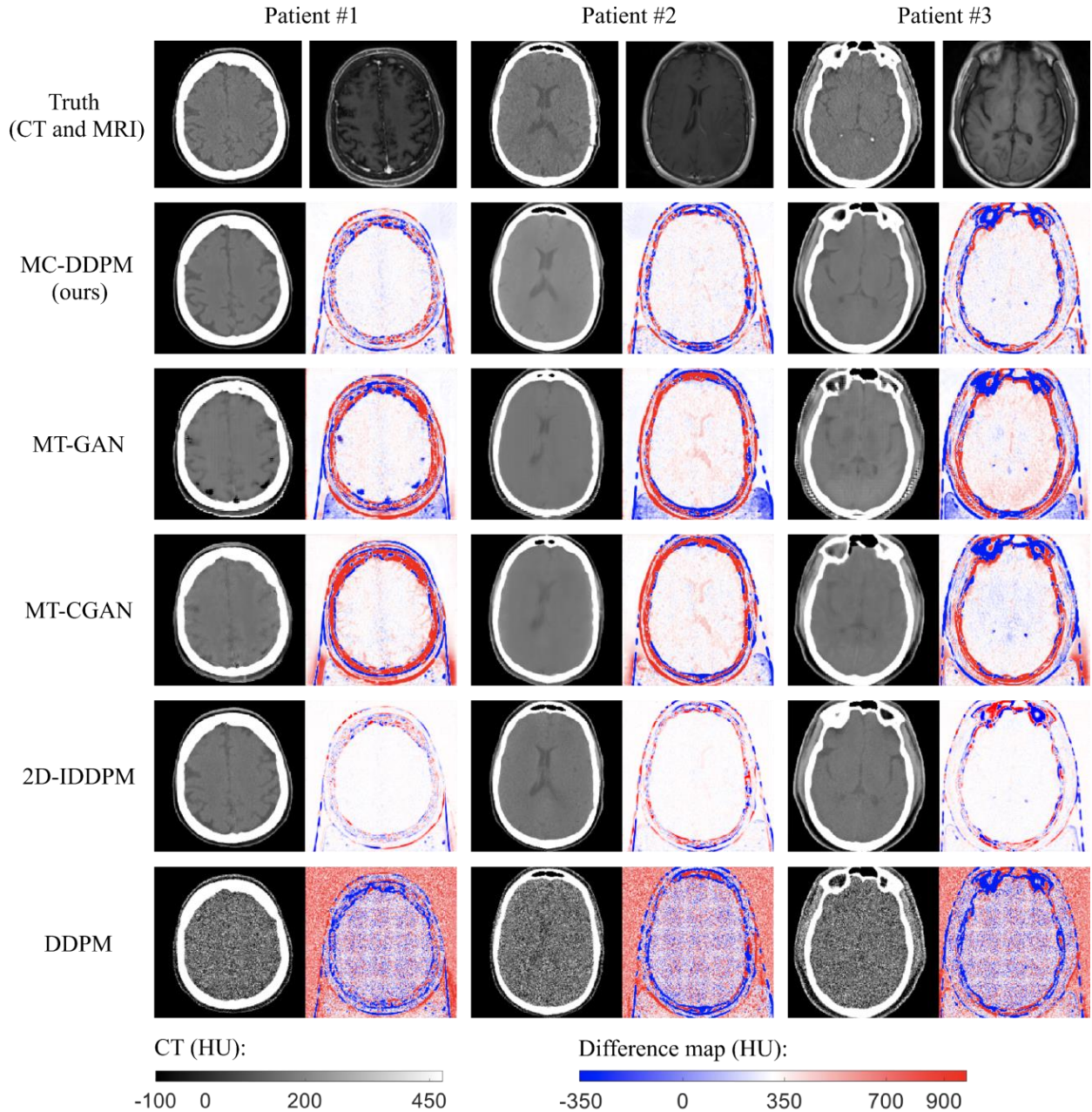

**Figure 3**: Synthetic CT images generated from the brain dataset including three ground truth slices from three patient subjects. The first row displays three ground truth CT images (left) and paired input MRIs (right). In the second to the sixth row, the sCT image outputs of MC-DDPM (row 2) and competing networks (row 3-6) are presented in the first, third and the fifth columns. The difference map between the sCTs with their corresponding ground truth CT images

are presented in the second, fourth, and the sixth column.

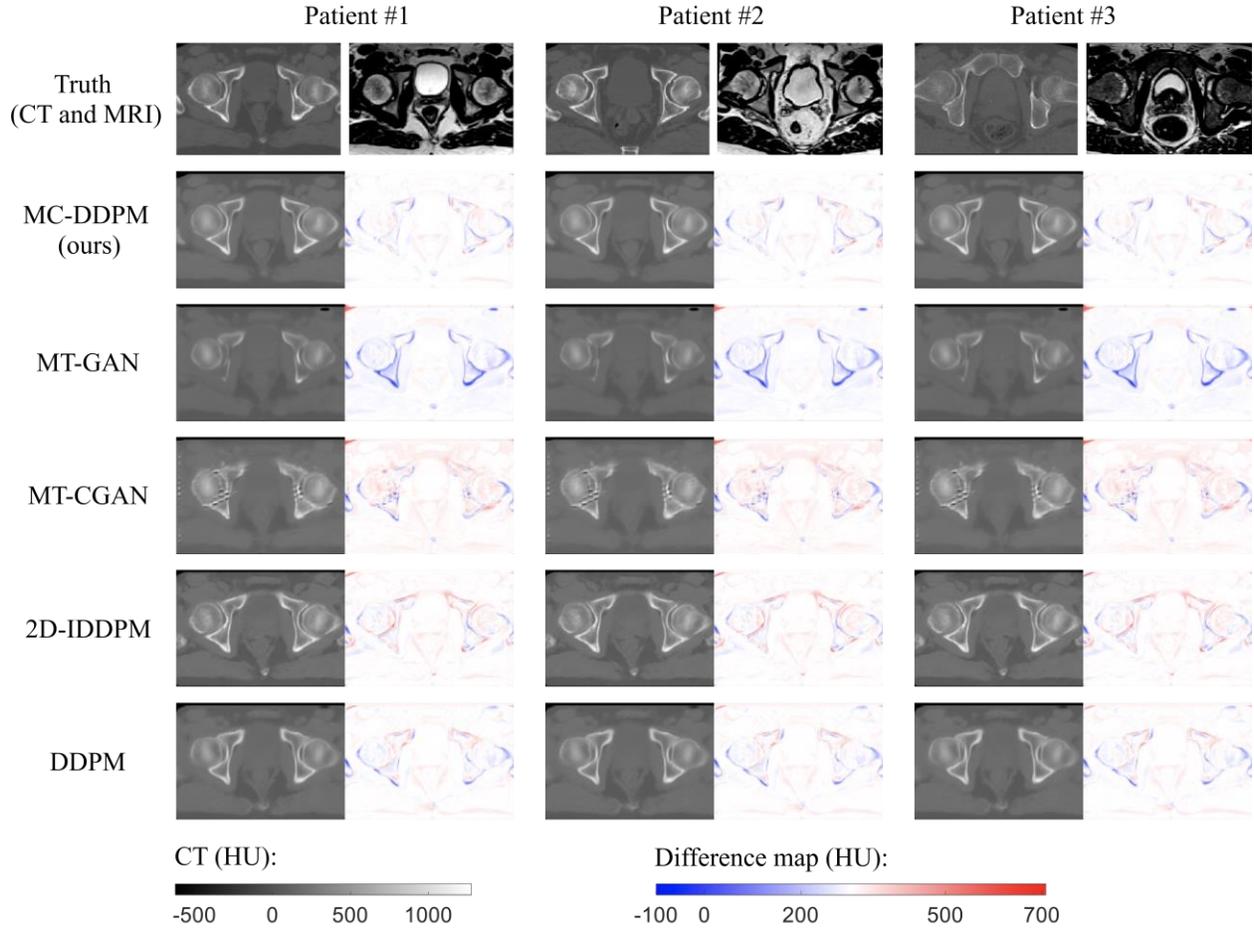

**Figure 4**: Synthetic CT images generated from the prostate dataset including three slices from three patient subjects. The first row displays three groups of the ground truth CT images and matched input MR scans. In the second to the sixth row, the sCT scans and the corresponding difference map of MC-DDPM (row 2) and competing networks (row 3-6) are presented in every two columns, respectively.

## 5. Results

sCTs generated using the brain and prostate datasets are displayed in Fig. 3 and 4. To quantitatively evaluate the performance of the MC-DDPM sCT synthesis from MRI, we present the quantitative and statistical comparison between MC-DDPM and other state-of-the-art methods.

### 5.A: Quantitative result using the brain dataset

Compared to all other methods, MC-DDPM has the lowest MAE (43.317±4.104 HU), indicating that it has the smallest average intensity difference between the predicted and ground truth images. Additionally, MC-DDPM has the highest PSNR (27.046±0.817 dB), highest SSIM (0.965±0.005) and NCC (0.983±0.004) among all techniques, indicating that it preserves the image structure and details better while providing the highest peak signal similarity, image visual similarity and overall image correlation. It should be noted that 2D-IDDPM exhibits superior quantitative results compared to GAN-based methods, while 3D-DDPM yields inferior results in contrast. One possible explanation is that DDPM was not originally designed to handle resampled generation timesteps: they require a large number of training and generation timesteps. Conversely, IDDPM was proposed specifically for the most efficient generation process (number of resampled timesteps can be much smaller than the number of training timesteps). Thus, it is possible that 3D-DDPM may produce worse results compared to GAN-based methods using the investigated settings. On the other hand, MC-DDPM achieves statistically significant improvements over all metrics when compared with MT-GAN, MT-CGAN, and 3D-DDPM. MC-DDPM also demonstrates significant improvements ($p<0.05$) when compared to 2D-IDDPM in terms of MAE, PSNR and NCC, but not SSIM.

## 5.B: Quantitative result using the prostate dataset

Compared to all other methods, MC-DDPM has the lowest MAE (59.953±12.462 HU), indicating that it produces prostate sCT images that have the smallest absolute difference relative to the ground truth CT images. Additionally, it has the highest PSNR (26.920±2.429 dB), SSIM (0.849±0.041) and NCC (0.948±0.018), indicating that the generated sCT images are closer to the peak signal ratio and visual appearance to ground truth, and have a higher correlation with the reference images. Similar to the brain sCT image evaluation, 2D-IDDPM outperforms GAN-based methods in terms of quantitative results, while 3D-DDPM shows worse results.

MC-DDPM demonstrates significant improvements over all metrics ($p <0.05$) when compared to GAN-based methods. Compared to 2D-IDDPM, MC-DDPM shows significant improvement ($p < 0.05$) in MAE and SSIM. Furthermore, compared to the 3D-DDPM, the MC-DDPM shows a statistically significant improvement (p-value $< 0.05$) across all metrics.

**Table 1.** Quantitative analysis of sCT images from MC-DDPM vs. MT-GAN, MT-CGAN, 2D-IDDPM, and 3D-DDPM using the institutional brain dataset. The table highlights the best-performing network(s), indicated in bold, and the second-best network(s), underlined, based on the mean evaluation results. P-values are provided to compare the results of MC-DDPM with those of the other competing methods. The reported values in the table are rounded to three decimal places. The unit of MAE is HU and the unit of PSNR is dB.

|  | MAE (↓) | PSNR (↑) | SSIM (↑) | NCC (↑) |
|---|---|---|---|---|
| MC-DDPM | **43.317±4.104** | **27.046±0.817** | **0.965±0.005** | **0.983±0.004** |
| *p-value* | N/A | N/A | N/A | N/A |
| MT-GAN | 69.967±8.673 | 23.360±0.869 | 0.916±0.010 | 0.960±0.009 |
| *p-value* | <0.010 | <0.010 | <0.010 | <0.010 |
| MT-CGAN | 66.139±5.221 | 24.101±0.688 | 0.936±0.008 | 0.967±0.007 |
| *p-value* | <0.010 | <0.010 | <0.010 | <0.010 |
| 2D-IDDPM | <u>50.512±3.670</u> | <u>25.899±0.692</u> | <u>0.963±0.004</u> | <u>0.977±0.005</u> |
| *p-value* | <0.010 | <0.010 | 0.055 | <0.010 |
| 3D-DDPM | 85.973±4.527 | 24.902±0.713 | 0.918±0.008 | 0.973±0.006 |
| *p-value* | <0.010 | <0.010 | <0.010 | <0.010 |

**Table 2.** Quantitative analysis of sCTs from MC-DDPM vs. MT-GAN, MT-CGAN, 2D-IDDPM, and 3D-DDPM using the prostate institutional dataset. The table highlights the best-performing network(s), indicated in bold, and the second-best network(s), underlined, based on the mean evaluation results. P-values are shown below each competing method. The unit of MAE is HU and the unit of PSNR is dB.

|  | MAE (↓) | PSNR (↑) | SSIM (↑) | NCC (↑) |
|---|---|---|---|---|
| MC-DDPM | **59.953±12.462** | **26.920±2.429** | **0.849±0.041** | **0.948±0.018** |
| *p-value* | N/A | N/A | N/A | N/A |
| MT-GAN | 85.794±18.496 | 24.507±2.366 | 0.789±0.043 | 0.916±0.033 |
| *p-value* | <0.010 | 0.019 | <0.010 | 0.051 |
| MT-CGAN | 75.147±9.674 | 25.786±1.586 | 0.831±0.039 | <u>0.940±0.019</u> |
| *p-value* | <0.010 | 0.042 | 0.044 | 0.031 |
| 2D-IDDPM | <u>68.336±12.074</u> | <u>25.841±2.088</u> | <u>0.840±0.038</u> | 0.938±0.021 |
| *p-value* | <0.010 | 0.059 | 0.034 | 0.108 |
| 3D-DDPM | 76.995±20.055 | 25.573±2.455 | 0.824±0.044 | 0.929±0.026 |
| *p-value* | 0.024 | 0.019 | 0.030 | 0.034 |

## 6. Discussion

We propose an innovative approach to generate high-quality sCT images from MRI using a novel algorithm: MC-DDPM. MC-DDPM utilizes diffusion models combined with Swin-transformer neural networks to produce accurate and reliable sCT images. The proposed method could eliminate the need for CT scans in radiation therapy planning, reducing radiation exposure and associated costs while improving patient comfort. The MC-DDPM consists of two main processes: forward diffusion and reverse diffusion processes. The forward diffusion process generates noisy images by adding Gaussian noise to the scanner-acquired CTs in multiple timesteps. This process generates a sequence of noisy CT scans with increasing noise levels, which are used as input for reverse diffusion. The reverse diffusion process iteratively denoises the sequence of noisy images using a proposed Swin-Vnet network. Specifically, the reverse diffusion process defines denoising as a prediction task: calculating a Gaussian Markov process's mean and variance. The Swin-Vnet network is trained to estimate the accumulated noise added at each timestep from a noisy CT scan with conditioning on an MR scan, which can be used for calculating the means and predicting a variance interpolation coefficient, which can be used for variance. An optimal Swin-Vnet allows the denoising process to iteratively transform pure Gaussian noise into a synthetic CT scan corresponding to the given MR scan. The Swin-Vnet follows a U-shape encoder-decoder architecture, where the encoder down-samples the input images to learn features in different resolution levels, and the decoder up-samples the features to the original input size to estimate the noise and the variance interpolation coefficient. Convolutional layers are deployed in high-resolution levels to capture local information efficiently. In contrast, Swin-attentions layers, which take advantage of Swin self-attention mechanisms, are set to low-resolution levels to model the global-level characteristics and refine the network's performance. This framework is the first diffusion model for sCT synthesis from MRI and the first attempt to utilize a 3D Swin-transformer-based network in designing a diffusion model to enhance the image synthesis quality.

In the institutional brain and prostate dataset, the MC-DDPM achieves state-of-the-art results: 1) In the brain dataset, by average among all the testing patients, the methods can generate sCT achieving MAE 43.317±4.104 HU, PSNR 27.046±0.817 dB, SSIM 0.965±0.005, and NCC 0.983±0.004. MC-DDPM demonstrates statistical improvement ($p < 0.05$) over the competing methods. 2) In the prostate dataset, the MC-DDPM can generate sCT achieving MAE 59.953±12.462 HU, PSNR 26.920±2.429 dB, SSIM 0.849±0.041, and NCC 0.948±0.018. MC-DDPM can obtain statistical improvements over all metrics compared to the MT-GAN, MT-CGAN, and 3D-DDPM, and improvement over the MAE and SSIM compared to the 2D-IDDPM. MC-DDPM therefore demonstrates utility in generating synthetic CT (sCT) images from MRIs, effectively streamlining the treatment planning workflow, reducing inefficiency, and potentially improving the patient experience while reducing costs.

GAN-based methods are sensitive to hyper-parameter settings and fine-tuning them requires significant time and computing resources for each dataset in order to produce acceptable outputs. In contrast, the MC-DDPM and DDPM proposed here are much less sensitive and exhibit greater stability. From the presented results, we can infer that MC-DDPM might easily be implemented for 3D translation tasks across other modalities, such as T1 or T2-weighted MRI or ultrasound, which we plan to investigate in a future study.

The exceptional performance of MC-DPPM comes at the cost of computational efficiency.

It took approximately 298 seconds for MC-DDPM using a sampling timestep of 50 and MC-based generation of only a single run to generate a 256x256x32 resolution sCT on the workstation described in Section 4. PGAN and C-PGAN were able to generate the same resolution image in only 22 seconds. The proposed diffusion model demonstrates high image quality, while low efficiency: First, a high quality sCT requires a relatively large number of generation iterations. Second, the stochasticity of generation requires MC-based generation which contains multiple runs for a single image. These limitations make 3D MRI to sCT synthesis prohibitively slow, and it can be inferred that the same low efficiency will be observed for other 3D images. Nevertheless, this limitation does not detract from the value of MC-DDPM, but instead highlights the need for further hardware and software optimization. Song *et al.*[37], Zhang *et al.*[38] and Kong *et al.*[39] have proposed algorithms for improving efficiency, such as designing an exponential forward process to generate noisy images with fewer timesteps or deploying a pre-trained network to accelerate the image synthesis process. These methods have the potential to enhance the efficiency of MC-DDPM. In addition, 3D diffusion-based methods for ultrasound and cone beam CT [1] have not yet been explored and evaluating the impact of synthetic medical images on these applications[40-42] is another promising area of future inquiry.

We further intend to explore efficiency improvements for the diffusion framework for 3D synthesis, investigating more sophisticated network architectures to elevate the quality of image synthesis, and designing a deterministic diffusion process to eliminate the need for multiple runs of Monte Carlo-based generation. Moreover, we plan to extend the application of MC-DDPM to a wider range of medical image modalities and undertake a more comprehensive study to validate the effectiveness of our approach.

## 7. Conclusion

This work presents a 3D magnetic resonance image to computed tomography image (MRI-to-CT) denoising diffusion probabilistic model (MC-DDPM) for generating synthetic CT scans from MR scans. The proposed method utilizes a 3D Shifted-window (Swin) transformer network to learn a diffusion process to convert a pure Gaussian noise into a realistic CT scan from a given MR scan. The method can achieve superior image quality compared to several competing state-of-the-art synthesis algorithms (GANs and conventional DDPMs). MC-DDPM generates high quality sCT images using only MRI inputs, eliminating the need for CT scans in radiation therapy planning, therefore improving quality of patient care.

**Acknowledgement**

This research was supported in part by National Institutes of Health R01CA215718, R56EB033332, R01EB032680 and R01CA272991.

**Reference:**

1. Peng J, Qiu RL, Wynne JF, et al. CBCT-Based Synthetic CT Image Generation Using Conditional Denoising Diffusion Probabilistic Model. *arXiv preprint arXiv:230302649.* 2023.
2. Chang C-W, Gao Y, Wang T, et al. Dual-energy CT based mass density and relative stopping power estimation for proton therapy using physics-informed deep learning. *Physics in Medicine & Biology.* 2022;67(11):115010.
3. Zhao B, Cheng T, Zhang X, et al. CT synthesis from MR in the pelvic area using Residual Transformer Conditional GAN. *Computerized Medical Imaging and Graphics.* 2023;103:102150.
4. Andreasen D, Van Leemput K, Edmund JM. A patch-based pseudo-CT approach for MRI-only radiotherapy in the pelvis. *Medical Physics.* 2016;43(8Part1):4742-4752.
5. Chowdhury N, Toth R, Chappelow J, et al. Concurrent segmentation of the prostate on MRI and CT via linked statistical shape models for radiotherapy planning. *Medical Physics.* 2012;39(4):2214-2228.
6. Olberg S, Zhang H, Kennedy WR, et al. Synthetic CT reconstruction using a deep spatial pyramid convolutional framework for MR-only breast radiotherapy. *Medical Physics.* 2019;46(9):4135-4147.
7. Zhao S, Geng C, Guo C, Tian F, Tang X. SARU: A self-attention ResUNet to generate synthetic CT images for MR-only BNCT treatment planning. *Medical Physics.* 2023;50(1):117-127.
8. Demol B, Boydev C, Korhonen J, Reynaert N. Dosimetric characterization of MRI-only treatment planning for brain tumors in atlas-based pseudo-CT images generated from standard T1-weighted MR images. *Medical Physics.* 2016;43(12):6557-6568.
9. Lei Y, Harms J, Wang T, et al. MRI-based synthetic CT generation using semantic random forest with iterative refinement. *Physics in Medicine & Biology.* 2019;64.
10. Yang H, Sun J, Carass A, et al. Unpaired brain MR-to-CT synthesis using a structure-constrained CycleGAN. Paper presented at: Deep Learning in Medical Image Analysis and Multimodal Learning for Clinical Decision Support: 4th International Workshop, DLMIA 2018, and 8th International Workshop, ML-CDS 2018, Held in Conjunction with MICCAI 2018, Granada, Spain, September 20, 2018, Proceedings 42018.
11. Li Y, Wang J, Chang C-W, et al. Multi-Parametric MRI radiomics analysis with ensemble learning for prostate lesion classification. Paper presented at: Medical Imaging 2023: Biomedical Applications in Molecular, Structural, and Functional Imaging2023.
12. Li Y, Wynne J, Wang J, et al. Cross-Shaped Windows Transformer with Self-supervised Pretraining for Clinically Significant Prostate Cancer Detection in Bi-parametric MRI. *arXiv preprint arXiv:230500385.* 2023.
13. Jonsson J, Karlsson M, Karlsson M, Nyholm T. Treatment planning using MRI data: an analysis of the dose calculation accuracy for different treatment regions. *Radiation Oncology (London, England).* 2010;5:62 - 62.
14. Hu M, Wang J, Chang C-W, Liu T, Yang X. *End-to-end brain tumor detection using a graph-feature-based classifier.* Vol 12468: SPIE; 2023.
15. Pan S, Chang C-W, Wang T, et al. Abdomen CT multi-organ segmentation using token-based MLP-Mixer. *Medical Physics.*n/a(n/a).
16. He K, Zhang X, Ren S, Sun J. Deep Residual Learning for Image Recognition. Paper presented at: 2016 IEEE Conference on Computer Vision and Pattern Recognition (CVPR); 27-30 June 2016, 2016.
17. Goodfellow I, Pouget-Abadie J, Mirza M, et al. Generative adversarial networks. *Communications of the ACM.* 2020;63(11):139-144.
18. Wolterink JM, Dinkla AM, Savenije MH, Seevinck PR, van den Berg CA, Išgum I. Deep MR to CT synthesis using unpaired data. Paper presented at: Simulation and Synthesis in Medical Imaging: Second International Workshop, SASHIMI 2017, Held in Conjunction with MICCAI 2017, Québec City, QC, Canada, September 10, 2017, Proceedings 22017.

19. Lei Y, Harms J, Wang T, et al. MRI-only based synthetic CT generation using dense cycle consistent generative adversarial networks. *Medical physics.* 2019.
20. Pan S, Flores J, Lin CT, Stayman JW, Gang GJ. Generative adversarial networks and radiomics supervision for lung lesion synthesis. Paper presented at: Medical Imaging 2021: Physics of Medical Imaging2021.
21. Shokraei Fard A, Reutens DC, Vegh V. From CNNs to GANs for cross-modality medical image estimation. *Computers in Biology and Medicine.* 2022;146:105556.
22. Song J, Meng C, Ermon S. Denoising diffusion implicit models. *arXiv preprint arXiv:201002502.* 2020.
23. Ho J, Jain A, Abbeel P. Denoising diffusion probabilistic models. *Advances in Neural Information Processing Systems.* 2020;33:6840-6851.
24. Dhariwal P, Nichol A. Diffusion models beat gans on image synthesis. *Advances in Neural Information Processing Systems.* 2021;34:8780-8794.
25. Nichol AQ, Dhariwal P. Improved Denoising Diffusion Probabilistic Models. Proceedings of the 38th International Conference on Machine Learning; 2021; Proceedings of Machine Learning Research.
26. Song Y, Sohl-Dickstein J, Kingma DP, Kumar A, Ermon S, Poole B. Score-based generative modeling through stochastic differential equations. *arXiv preprint arXiv:201113456.* 2020.
27. Gui J, Sun Z, Wen Y, Tao D, Ye J. A review on generative adversarial networks: Algorithms, theory, and applications. *IEEE Transactions on Knowledge and Data Engineering.* 2021.
28. Wolleb J, Bieder F, Sandkühler R, Cattin PC. Diffusion Models for Medical Anomaly Detection. *arXiv preprint arXiv:220304306.* 2022.
29. Pan S, Wang T, Qiu RL, et al. 2D medical image synthesis using transformer-based denoising diffusion probabilistic model. *Physics in Medicine and Biology.* 2023.
30. Lyu Q, Wang G. Conversion Between CT and MRI Images Using Diffusion and Score-Matching Models. *arXiv preprint arXiv:220912104.* 2022.
31. Pan S, Chang C-W, Peng J, et al. Cycle-guided Denoising Diffusion Probability Model for 3D Cross-modality MRI Synthesis. *arXiv preprint arXiv:230500042.* 2023.
32. He K, Zhang X, Ren S, Sun J. Deep residual learning for image recognition. Paper presented at: Proceedings of the IEEE conference on computer vision and pattern recognition2016.
33. Wu Y, He K. Group normalization. Paper presented at: Proceedings of the European conference on computer vision (ECCV)2018.
34. Elfwing S, Uchibe E, Doya K. Sigmoid-weighted linear units for neural network function approximation in reinforcement learning. *Neural Networks.* 2018;107:3-11.
35. Ho J, Saharia C, Chan W, Fleet DJ, Norouzi M, Salimans T. Cascaded Diffusion Models for High Fidelity Image Generation. *J Mach Learn Res.* 2022;23(47):1-33.
36. Loshchilov I, Hutter F. Decoupled weight decay regularization. *arXiv preprint arXiv:171105101.* 2017.
37. Song Y, Dhariwal P, Chen M, Sutskever I. Consistency models. *arXiv preprint arXiv:230301469.* 2023.
38. Zhang Q, Chen Y. Fast Sampling of Diffusion Models with Exponential Integrator. *arXiv preprint arXiv:220413902.* 2022.
39. Kong Z, Ping W. On fast sampling of diffusion probabilistic models. *arXiv preprint arXiv:210600132.* 2021.
40. Hu M, Wang J, Chang C-w, Liu T, Yang X. *Ultrasound breast tumor detection based on vision graph neural network.* Vol 12470: SPIE; 2023.
41. Hu M, Wang J, Wynne J, Liu T, Yang X. *A vision-GNN framework for retinopathy classification using optical coherence tomography.* Vol 12465: SPIE; 2023.
42. Li Y, Zhou B, Wang J, et al. Ultrasound-based dominant intraprostatic lesion classification with Swin Transformer. Paper presented at: Medical Imaging 2023: Ultrasonic Imaging and Tomography2023.